\documentclass[%
 reprint,
 superscriptaddress,
 amsmath,amssymb,
 aps,
]{revtex4-1}

\usepackage{graphicx}
\usepackage{dcolumn}
\usepackage{bm}
\usepackage{color}
\usepackage{amssymb}

\begin{document}

\preprint{APS/123-QED}

\title{Improving the performance of Twin-Field Quantum Key Distribution}
	\author{Feng-Yu Lu}
	\author{Zhen-Qiang Yin}
	\email{yinzq@ustc.edu.cn}
	\author{Chao-Han Cui}
    \author{Guan-Jie Fan-Yuan}
	\author{Rong Wang}
	\author{Shuang Wang}
	\author{Wei Chen}
	\author{De-Yong He}
	\affiliation{Key Laboratory of Quantum Information, CAS Center For Excellence in Quantum Information and Quantum Physics,
		University of Science and Technology of China, Hefei 230026, China}
	\affiliation{State Key Laboratory of Cryptology, P. O. Box 5159, Beijing 100878, P. R. China}
	\author{Wei Huang}
	\author{Bing-Jie Xu}
	\affiliation{Science and Technology on Communication Security Laboratory, Institute of Southwestern Communication, Chengdu, Sichuan 610041, China}
	\author{Guang-Can Guo}
	\author{Zheng-Fu Han}
	\affiliation{Key Laboratory of Quantum Information, CAS Center For Excellence in Quantum Information and Quantum Physics,
		University of Science and Technology of China, Hefei 230026, China}
	\affiliation{State Key Laboratory of Cryptology, P. O. Box 5159, Beijing 100878, P. R. China}




\date{\today}

\begin{abstract}
Among the various versions of the twin-field quantum key distribution (TF-QKD) protocol [M.Lucamarini, Z. Yuan, J. Dynes, and A. Shields, Nature (London) 557, 400 (2018)] that can overcome the rate-distance limit, the TF-QKD without phase postselection proposed by Cui et al. [Phys. Rev. Appl. 11, 034053 (2019)] is an elegant TF-QKD that can provide high key rates since the postselection of global phases has been removed. However, the achievable distance of this variant is shorter than that of the original phase-matching QKD [X. Ma, P. Zeng, and H. Zhou, Phys. Rev. X 8, 031043 (2018)]. In this paper, we propose a method for improving its performance by introducing an additional decoy mode. The upper bound of the information leakage can be more tightly estimated; hence, both the key rate and the achievable distance are significantly improved. Interestingly, the operation of the proposed additional decoy mode is the same as that of the code mode; hence, it does not introduce difficulties into the experimental system. In addition, the improvement is substantial with finite decoy states, which is meaningful in practice.

\end{abstract}

\pacs{Valid PACS appear here}
\maketitle


\section{introduction}


Quantum key distribution (QKD) \cite{BB84} enables two remote users, who we call Alice and Bob, to share secret random keys with information-theoretic security \cite{lo1999unconditional,shor2000simple,Scarani:QKDrev:2009,Rennersecurity} that is guaranteed by principles of quantum physics, even if there is an eavesdropper, who we call Eve.

With the developments of QKDs in both theory and experiment, QKD implementations with longer achievable distance \cite{boaron2018secure,yin2016measurement} and higher secret key rate(SKR) \cite{wang20122,gordon2005quantum,wang2018practical} were realized.  However, all these implementations must obey limits on the SKR as a function of the channel transmittance \cite{takeoka2014fundamental,pirandola2017fundamental}, which are called repeaterless bounds.
Surprisingly, a recently proposed protocol, namely, twin-field QKD (TF-QKD) \cite{tfqkd} and its variants, e.g., phase-matching QKD(PM-QKD) \cite{pmqkd}, sending-or-not QKD \cite{sns-tfqkd} and no phase post-selection TF-QKD(NPP-TFQKD) \cite{cui2018phase,curty2018simple,lin2018simple}, can overcome this bound; hence, the performance of QKDs can be significantly improved. In addition, these protocols have been proven to be immune to all potential side-channel attacks on the measurement device, like the measurement-device-independent protocol \cite{lo2012measurement,ma2012alternative,liu2013experimental,wang2015phase,wang2017measurement}.

In the original TF-QKD and PM-QKD, Alice (Bob) encodes a key bit as the phase of the weak coherent pulse, adds an additional random phase $\alpha_A$ ($\alpha_B$), and sends it to an untrusted middle station, namely, Charlie, who interferes with the incoming pulses to measure the phase difference between them. Upon receiving the message from the middle station, Alice and Bob publicly announce the values of $\alpha_A$ and $\alpha_B$ and post-select the trials that satisfy $\alpha_A\approx \alpha_B$ to generate secret key bits. The post-selection of $\alpha_A\approx \alpha_B$ inevitably degrades the SKR and complicates the postprocessing. For overcoming this problem, Cui et al. proposed NPP-TFQKD \cite{cui2018phase}. Soon after, two other groups independently proposed similar schemes \cite{curty2018simple,lin2018simple}. In \cite{cui2018phase}, Alice(Bob) randomly selects two different modes. The first mode runs without adding random phase $\alpha_A$ ($\alpha_B$) and can be used to generate a key bit. The latter is used to monitor the security. Removing the phase randomization and the post-selection from the code mode observably improves the secure key rate. However, its achievable distance is much shorter than that of PM-QKD.

In this work, we explain why the achievable distance is shorter compared with PM-QKD and propose a practical method for improving the performance of NPP-TFQKD substantially. Our work is mainly based \cite{cui2018phase} and the core strategy of our method is to introduce an additional decoy mode into the NPP-TFQKD protocol that is run with the same phase as code mode and can estimate tightly the information leakage, which is denoted as $I_{AE}$. As a result, the achievable distance is increased.

The remainder of this paper is organized as follows: In Sec.\textrm{II}, we briefly review the procedure of NPP-TFQKD and its method for calculating the upper bound of $I_{AE}$ from \cite{cui2018phase}. In Sec.\textrm{III}, we introduce our new method, which simultaneously maintains the superiority of the higher SKR, longer distance and practicability. In Sec.\textrm{IV}, the simulation results are presented, which demonstrate the superior performance of our method. The details of our method and derivations can be found in the appendix.

\section{TF-QKD without phase post selection}

The process of NPP-TFQKD is described as follows:


\textbf{Step 1. Preparation and measurement:} Alice and Bob randomly select code mode or decoy mode. If code mode is selected, Alice (Bob) prepares a phase-locked weak coherent pulse (WCP) $|\sqrt{\mu}\rangle$ and randomly modulates the $0$ or $\pi$ phase that corresponds to the raw key bit $0$ or $1$, respectively. If the decoy mode is selected, they prepare a phase-randomized WCP with an intensity that is selected from a pre-specified set at random. Then, they send the modulated quantum state to Charlie for interference. Since the randomized phase in the decoy mode is never publicly announced, we assume Alice and Bob prepare a mixed state in photon number space assume
\begin{equation}
\rho_{\omega_1}\otimes\rho_{\omega_2} = \sum_{m=0}^\infty \sum_{n=0}^\infty (p_m^{\omega_1} |{m}\rangle\langle{m}|)\otimes(p_n^{\omega_2}|{n}\rangle\langle{n}|),
\label{WCSstate}
\end{equation}
where $\omega_1$ and $\omega_2$ denote, respectively, the intensity chosen by Alice and Bob, and $p_n^{\omega}=e^{-\omega}\omega^n/n!$.

\textbf{Step 2. Announcement:} For each trial, Charlie must publicly announce the detector (L or R) that clicks or a non-click event.

\textbf{Step 3. Sifting:} Alice and Bob repeat the above steps many times to collect sufficiently many click events. Then, they publicly announce which trials are conducted in code mode and which are conducted in decoy mode. For the trials, they both choose code mode and Charlie announces a click event (L or R clicked); the raw key bits are generated. Bob must flip his bit if Charlie's announcement is 'R'.

\textbf{Step 4. Parameter estimation:} Alice and Bob estimate the gain of code mode $Q_{\mu\mu}^c$, namely, the probability of a click event for each trial in code mode, and the gains in decoy mode. For decoy mode, we denote the gains as $Q_{\omega_1 \omega_2}^d$, which corresponds to the probability of a click event conditioned on Alice and Bob preparing $\rho_{\omega_1}\otimes\rho_{\omega_2}$. In decoy mode, we also define the yield of Fock states $(|{m}\rangle\langle{m}|\otimes|{n}\rangle\langle{n}|)$ as $Y_{m,n}$. The relation between them is expressed as
\begin{equation}
Q_{\omega_1 \omega_2} = \sum_{m=0}^\infty \sum_{n=0}^\infty p_m^{\omega_1} p_n^{\omega_2} Y_{m,n}
\label{eq_Q_Y},
\end{equation}
where $Q_{\omega_1 \omega_2}$ and $p^{\omega_1(\omega_2)}_{n(m)}$ can be experimentally observed.

From these gains, Alice and Bob can calculate the upper bound on the information leakage, which is denoted as $\overline{I}_{AE}$.

\textbf{Step 5. Key distillation:} Alice and Bob generate their secure key by applying error correction and privacy amplification to the sifted key. The SKR is expressed as
\begin{equation}
R = Q^c_{\mu\mu}[1 - fH(E^c) - \overline{I}_{AE}],
\label{SKR}
\end{equation}
where $H(p) = -p\text{log}_2p - (1-p)\text{log}_2(1-p)$ is the Shannon entropy.

In \cite{cui2018phase}, $\overline{I}_{AE}$ depends on $|\gamma_{n,m} \rangle$, which is defined as the state of Eve's ancilla if Alice and Bob send Fock states $|n\rangle$ and $|m\rangle$, respectively. Defining $\mathbb{E}=\{0,2,4,....\}$ and $\mathbb{O}=\{1,3,5,....\}$ as even and odd sets, respectively, the relation between $I_{AE}$ and $|\gamma_{n,m} \rangle$ is as follows:
	
\begin{equation}
\label{I_AE}
I_{AE} \le h(\dfrac{\big||\psi_{ee}\rangle\big|^2}{Q_{\mu\mu}^c},\dfrac{\big||\psi_{oe}\rangle\big|^2}{Q_{\mu\mu}^c})+h(\dfrac{\big||\psi_{oo}\rangle\big|^2}{Q_{\mu\mu}^c},\dfrac{\big||\psi_{eo}\rangle\big|^2}{Q_{\mu\mu}^c}),
\end{equation}
where $h(x,y)=-x\log_2x-y\log_2y+(x+y)\log_2 (x+y)$ and
\begin{equation}
\label{intermediateState}
\begin{aligned}
&|\psi_{ee}\rangle  = \sum_{n\in{\mathbb{E}},m\in{\mathbb{E}}}{\sqrt{p^\mu_{n}p^\mu_{m}Y_{n,m}}}|\gamma_{n,m} \rangle, \\
&|\psi_{oe}\rangle  = \sum_{n\in{\mathbb{O}},m\in{\mathbb{E}}}{\sqrt{p^\mu_{n}p^\mu_{m}Y_{n,m}}}|\gamma_{n,m} \rangle, \\
&|\psi_{oo}\rangle  = \sum_{n\in{\mathbb{O}},m\in{\mathbb{O}}}{\sqrt{p^\mu_{n}p^\mu_{m}Y_{n,m}}}|\gamma_{n,m} \rangle,\\
&|\psi_{eo}\rangle  = \sum_{n\in{\mathbb{E}},m\in{\mathbb{O}}}{\sqrt{p^\mu_{n}p^\mu_{m}Y_{n,m}}}|\gamma_{n,m} \rangle.
\end{aligned}
\end{equation}
The subscript $e$ ($o$) denotes that the photon-number $n$ or $m$ belongs to set $\mathbb{E}$ ($\mathbb{O}$).

 The main challenging is bounding $\big| |\psi_{xy}\rangle\big|^2(x,y\in\{o,e\})$.
Since the values of inner product $\langle\gamma_{n,m}|\gamma_{k,l} \rangle(m,n\neq k,l)$ are unknown, what we can make sure are the constraints given by:

\begin{equation}
\begin{aligned}
\label{cons}
&x_{ee}\triangleq\big||\psi_{ee}\rangle\big|^2\leqslant\big|\sum_{n\in{\mathbb{E}},m\in{\mathbb{E}}}\sqrt{p^\mu_{n}p^\mu_{m}Y_{n,m}} \big|^2, \\
&x_{oe}\triangleq\big||\psi_{oe}\rangle\big|^2\leqslant\big|\sum_{n\in{\mathbb{O}},m\in{\mathbb{E}}}\sqrt{p^\mu_{n}p^\mu_{m}Y_{n,m}} \big|^2,\\
&x_{oo}\triangleq\big||\psi_{oo}\rangle\big|^2\leqslant\big|\sum_{n\in{\mathbb{O}},m\in{\mathbb{O}}}\sqrt{p^\mu_{n}p^\mu_{m}Y_{n,m}} \big|^2,\\
&x_{eo}\triangleq\big||\psi_{eo}\rangle\big|^2\leqslant\big|\sum_{n\in{\mathbb{E}},m\in{\mathbb{O}}}\sqrt{p^\mu_{n}p^\mu_{m}Y_{n,m}} \big|^2,\\
&x_{ee}+x_{oe}+x_{oo}+x_{eo}=Q_{\mu\mu}^c.
\end{aligned}
\end{equation}

The upper bound $\overline{I}_{AE}$ can be obtained by maximizing the objective equation
\begin{equation}
\begin{aligned}
\overline{I}_\text{AE}
=\text{max}:\,\,  h(\dfrac{x_{ee}}{Q_{\mu\mu}^c},\dfrac{x_{oe}}{Q_{\mu\mu}^c})+h(\dfrac{x_{oo}}{Q_{\mu\mu}^c},\dfrac{x_{eo}}{Q_{\mu\mu}^c})
\label{I_EA_upper}
\end{aligned}
\end{equation}
without violating the constraints in Eq.(\ref{cons}).

A very interesting and meaningful question is whether the above bound on $I_{AE}$ can be tightened.

\section{Method to improve the distance of NPP-TFQKD}


According to the constraints in Eq.(\ref{cons}), we posit that $I_{AE}$ is bounded loosely. Let's consider the upper bound of $\big||\psi_{ee}\rangle\big|^2$ as an example. Defining $q_{n,m} =  p^\mu_{n}p^\mu_{m}Y_{n,m}$, $\big||\psi_{ee}\rangle\big|^2$ can be regarded as two parts: 

\begin{equation}
\label{Eq_psi}
\big||\psi_{ee}\rangle \big|^2 = \big|\sum_{n,m}^{n\in{\mathbb{E}},m\in{\mathbb{E}}}{\sqrt{q_{n,m}}}|\gamma_{n,m} \rangle\big|^2= \Omega_{ee}^{\mu} + \Phi_{ee}^{\mu},
\end{equation}
where
\begin{equation}
\label{omega_ee}
\begin{aligned}
\Omega_{ee}^{\mu\mu}& = \sum_{n,m}^{n\in{\mathbb{E}},m\in{\mathbb{E}}}q_{n,m}\langle\gamma_{n,m}|\gamma_{n,m}\rangle = \sum_{n,m}^{n\in{\mathbb{E}},m\in{\mathbb{E}}}q_{n,m}
\end{aligned}
\end{equation}
represents sum of the inner products whose subscripts of bra and ket are the same, 
and
\begin{equation}
\label{phi_ee}
\Phi_{ee}^{\mu\mu}=\sum_{n,m,k,l>m}^{n,k\in{\mathbb{E}},m,l\in{\mathbb{E}}}\sqrt{q_{n,m}q_{k,l}}(\langle\gamma_{n,m}|\gamma_{k,l} \rangle + \langle\gamma_{k,l}|\gamma_{n,m} \rangle)
\end{equation} 
represents the remaining inner products, whose subscripts of bra and ket are different. We refer to $\Omega_{ee}$ and $\Phi_{ee}$ as the non-cross term and the cross term, respectively.

Since $\langle\gamma_{n,m}|\gamma_{k,l} \rangle + \langle\gamma_{k,l}|\gamma_{n,m} \rangle = 2Re(\langle\gamma_{n,m}|\gamma_{k,l} \rangle)$ must be a real number in the range of $[-2,2]$, the original upper bound is estimated too loosely since Eq.(\ref{cons}) replaces all $\langle\gamma_{n,m}|\gamma_{k,l} \rangle + \langle\gamma_{k,l}|\gamma_{n,m} \rangle$ by $2$, which is the worst case. However, if Charlie is honest, his interference measurement preserves the orthogonality between Fock states, from which $\langle\gamma_{n,m}|\gamma_{k,l} \rangle + \langle\gamma_{k,l}|\gamma_{n,m} \rangle=0$ follows directly. Thus, this replacement may severely degrade SKR if the channel loss is large, which is the main reason why the distance of NPP-TFQKD is shorter than that of PM-QKD. It is natural to consider whether there is any method for estimating $\overline{I}_{AE}$ more tightly while preserving the practicability of the protocol. Fortunately, the answer is yes.

By observing Eq.(\ref{eq_Q_Y}) and the fifth constraint condition in Eq.(\ref{cons}), we can find the gains of code mode and decoy mode are very different, since the phase randomization in decoy mode eliminates all cross terms. Concretely, we can see

\begin{equation}
\label{compare_Q}
\begin{aligned}
&Q^{d}_{\omega_1 \omega_2 } =\Omega_{ee}^{\omega_1 \omega_2 }+\Omega_{eo}^{\omega_1 \omega_2 }+\Omega_{oe}^{\omega_1 \omega_2 }+\Omega_{oo}^{\omega_1 \omega_2 } \\
&Q^{c}_{\omega_1 \omega_2 } = Q^{d}_{\omega_1 \omega_2 }+\Phi_{ee}^{\omega_1 \omega_2 }+\Phi_{eo}^{\omega_1 \omega_2 }+\Phi_{oe}^{\omega_1 \omega_2 }+\Phi_{oo}^{\omega_1 \omega_2 }
\end{aligned}
\end{equation}

By defining
\begin{equation}
y_{n,m,k,l} = (\langle\gamma_{n,m}|\gamma_{k,l} \rangle + \langle\gamma_{k,l}|\gamma_{n,m} \rangle)\sqrt{Y_{n,m}Y_{k,l}},
\end{equation}
we obtain a new linear equation in terms of $y_{n,m,k,l}$:


\begin{equation}
\label{new_equation}
\begin{aligned}
&Q_{\omega_1 \omega_2}^{c} - Q_{\omega_1 \omega_2}^{d} 
=\sum_{\mathbb{A},\mathbb{B}}^{\mathbb{A},\mathbb{B}\in{\{\mathbb{O},\mathbb{E}\}}}\sum_{n,k\in{\mathbb{A}},m,l\in{\mathbb{B}}}^{l>m}\sqrt{{p^{\omega_1}_{n}p^{\omega_2}_{m}p^{\omega_1}_{k}p^{\omega_2}_{l}}}y_{n,m,k,l}\\
\end{aligned}
\end{equation}

Similar to the principle of the infinite decoy state method \cite{wang2005decoy,lo2005decoy,hwang2003quantum}, we can obtain infinite linear equations that are similar to Eq.(\ref{new_equation}) if infinite intensities in both code and decoy modes are applied. By solving these infinite linear equations, all $y_{n,m,k,l}$ can be obtained in principle. In the ideal scenario, in which Charlie is honest, $ Q_{\omega_1 \omega_2}^{c} - Q_{\omega_1 \omega_2}^{d} = 0$ should be satisfied for any intensities $\omega_1$ and $\omega_2$. To satisfy these infinite equations, $y_{n,m,k,l}=0$ must hold, namely, the cross term $\Phi_{ee}^{\omega_1 \omega_2}$ in Eq.(\ref{Eq_psi}) must be zero. As a result, tighter estimation of $I_{AE}$ and a higher SKR are expected.

To introduce the linear equations such as Eq.(\ref{new_equation}) into the NPP-TFQKD protocol, one must monitor the gains of various intensities with the same phase in code mode. For this and to avoid ambiguity, we modify \textbf{Step 1} and \textbf{Step 4} of NPP-TFQKD as follows:

To introduce the linear equations like Eq.(\ref{new_equation}) in NPP-TFQKD protocol, one must monitor the gain of various intensities with the same phase of code mode. For this and avoiding ambiguity, we modify the \textbf{Step.1} and \textbf{Step.4} of NPP-TFQKD as follows.
\\

\textbf{New Step 1. Preparation and measurement}: Alice and Bob randomly choose \textbf{decoy mode 1}, \textbf{decoy mode 2} or \textbf{code mode}. If \textbf{code mode} is selected, Alice (Bob) prepares a phase-locked WCP $|\sqrt{\mu}\rangle$ and randomly modulates the $0$ or $\pi$ phase that corresponds to raw key bit $0$ or $1$, respectively. If \textbf{decoy mode 1} is selected, they prepare a phase-randomized WCP and randomly choose an intensity from a pre-specified set $I_1$. If \textbf{decoy mode 2} is selected, Alice (Bob) prepares a phase-locked WCP $|\sqrt{\omega}\rangle$, in which the intensity $\omega$ is chosen from a pre-specified set $I_2$ at random. The quantum state of decoy mode 2 shares the same phase as WCP in code mode and $I_2$ is a subset of $I_1$.
\\

\textbf{New Step 4. Parameter estimation:} Alice and Bob estimate the gain of code mode $Q_{\mu\mu}^c$, namely, the probability of a click event for each trial in code mode, and the gains of decoy modes 1 and 2. For decoy mode 1, we denote the gain as $Q_{\omega_1 \omega_2}^{d_1}$, which corresponds to the probability of a click event conditioned on Alice and Bob preparing $\rho_{\omega_1}\otimes\rho_{\omega_2}$ in decoy mode 1. For decoy mode 2, we denote the gain as $Q_{\omega_1 \omega_2}^{d_2}$, which corresponds to the probability of a click event conditioned on Alice and Bob preparing $|\sqrt{\omega_1}\rangle$ and $|\sqrt{\omega_2}\rangle$, respectively, in decoy mode 2. From these gains, Alice and Bob can calculate the upper bound on the information leakage $\overline{I}_{AE}$. The method of calculating $I_{AE}$ will be detailed in the appendix. 

With the new step 1 and step 4, we have proposed an improved NPP-TFQKD.

\section {Simulation of the improved NPP-TFQKD}
\subsection{Infinite decoy states}

\begin{figure}[htbp]
 \includegraphics[width=9cm]{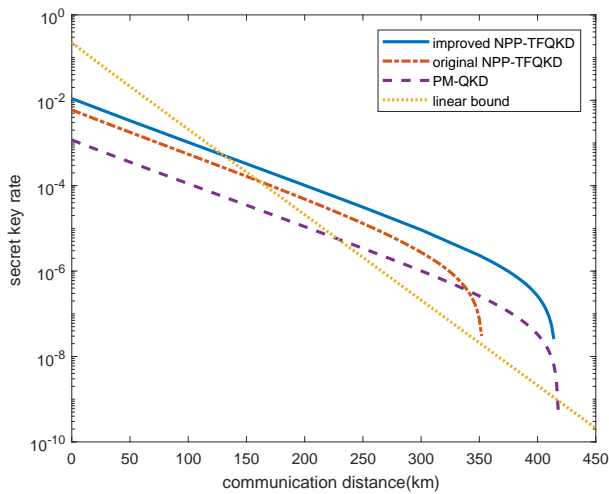}
\caption{\label{Fig_ideal} Secret key rates in logarithmic scale as functions of the distance for various protocols with infinite decoy states. The blue solid line and red dash-dot line correspond to, respectively, our improved method and the original NPP-TFQKD. The purple dashed line corresponds to PM-QKD. The yellow dotted line corresponds to the repeaterless bound that was proposed in \cite{pirandola2017fundamental}; here, we refer to it as the linear bound.}
\end{figure}

To evaluate the performance of our improved NPP-TFQKD, we simulate its SKR with infinite decoy states, namely, the sets $I_1$ and $I_2$ are both infinite; hence, $Y_{n,m}$ is calculated precisely and $y_{n,m,k,l}=0$.
The simulation model can be found in the appendix of Ref. \cite{cui2018phase} and the parameters that are used in the simulation are listed in Tab.(\ref{Tab_parameters}). The results of the simulation are plotted in Fig.(\ref{Fig_ideal}), according to which the proposed improved protocol realizes higher SKR and longer communication distance.

\begin{table}[b]
\caption{\label{Tab_parameters}
simulation parameters of several kinds of TF-QKD.}
\begin{ruledtabular}
\begin{tabular}{ccccc}
 $p_{dc}$\footnotemark[1] & $\eta_d$\footnotemark[2] & $f$\footnotemark[3]  & fiber loss & M\footnotemark[4] \\
\hline
$8\times{10^{-8}}$ & $14.5\%$ & $1.15$ & $0.2\ \text{dB/km}$ &16\\
\end{tabular}
\footnotetext[1]{$P_{dc}$ denotes the dark count rate}
\footnotetext[2]{$\eta_d$ denotes the detection efficiency}
\footnotetext[3]{$f$ denotes the correction efficiency}
\footnotetext[4]{$M$ denotes phase post selection slice number in PM-QKD}
\end{ruledtabular}
\end{table}

\subsection{finite decoy states}
The infinite decoy state method is not useful in practice. In this subsection, we evaluate the performance of applying 5-intensity decoy mode 1 and 4-intensity decoy mode 2.
The intensity of code mode is denoted by $\mu$. The 4 intensities of decoy mode 2 are $\mu$, $\nu_1$, $\nu_2$ and $o$. Decoy mode 1 has an additional intensity, which is denoted by $\mu_3$.
The parameters are listed in Tab.(\ref{Tab_parameters}). Compared with the original NPP-TFQKD with four decoy state intensities (red solid line), both the SKR and the achievable distance are improved substantially. The details will be introduced in the appendix.

\begin{figure}[htbp]
\includegraphics[width=9cm]{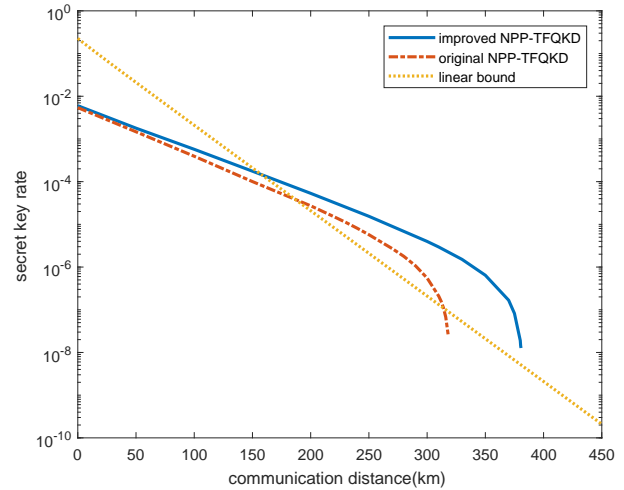}
\caption{\label{fig_practical} Secret key rates in logarithmic scale as functions of the distance for various practical protocols. The blue solid line corresponds to our improved method with 5-intensity decoy mode 1 and 4-intensity decoy mode 2. The red dash-dot line corresponds to the original NPP-TFQKD with 4 decoy states. The yellow dotted line represents the linear bound that was proposed in \cite{pirandola2017fundamental}.
The original NPP-TFQKD requires only code mode and decoy mode. The intensity $\mu$ (the first decoy state and the code state) is an optimized value. Other decoy intensities, namely, $\nu_1$, $\nu_2$ and $o$, are fixed to $0.005$, $0.002$ and $0$, respectively. For our improved protocol, the intensity $\mu$ is an optimized value. $\nu_1$, $\nu_2$ and $o$ are set to the same values as above. The additional intensity $\mu_3$ of decoy mode 1 is fixed to $1.3$.
}
\end{figure}


It is worth noting that, The estimation of cross terms relies on the accuracy of $\overline{Y}_{m,n}$. To estimate the high-order $\overline{Y}_{m,n}$($m+n\ge4$) more tightly, we add an intensity that is larger than the code mode intensity into decoy mode 1.

\section{Discussion}

During the submission of our work, we found that the test states in other work \cite{primaatmaja2019almost} are very similar to decoy mode 2. To help readers understand the variants of TF-QKD, we will briefly discuss Ref. \cite{primaatmaja2019almost,cui2018phase,curty2018simple,lin2018simple} and compare our work with ref. \cite{primaatmaja2019almost} with the same parameters and number of intensities.

The protocols in Ref. \cite{cui2018phase,curty2018simple} are highly similar; both of them have higher SKR but much shorter achievable distance compared with PM-QKD. In the similar scheme that was proposed by Ref. \cite{lin2018simple}, the achievable distance is increased by using infinite test states, which include infinite intensities and infinite phase modulation. The achievable distance is the same as that of PM-QKD but the infinite phase modulation for each intensity is not feasible in practice. Our method divides each of the intensities into only two modes: a phase-randomized mode, namely, decoy mode 1, and a phase-locked mode, namely, decoy mode 2; this approach is more feasible than infinite phase modulation. Our method also reaches the distance realized by PM-QKD.

In Ref. \cite{primaatmaja2019almost}, the finite test state method was proposed. The method renders the protocol in \cite{lin2018simple} useful in practice. Their method requires several test states for estimating the parameters and two key states for generating key bits (the two key states are also used for parameter estimation). We consider their six-test-state method as an example. The method requires six test states and two key states, which include four intensities and eight phase modulations in total. According to Fig.6(ii) of \cite{primaatmaja2019almost}, the protocol reaches 75 dB total loss with $5\times 10^{-8}$ dark-count rate, $85\%$ detection efficiency and $1.5\%$ misalignment. We simulate our protocol with the same parameters, 4-intensity decoy mode 1 and 4-intensity decoy mode 2. The results demonstrate that our protocol achieves a total loss of 94 dB.

Both this work and Ref. \cite{primaatmaja2019almost} have advantages and disadvantages.
For our protocol, the phase randomization in $[0,\pi]$ in the decoy mode renders the system more complicated and may limit the repetition rate. Nevertheless, the phase randomization in our protocol does not assume the precision of phase modulation if the phase is random. For the approach in \cite{primaatmaja2019almost}, one need only realize a small number of phase modulations; however, these phase modulations are assumed to be very precise. For instance, the six test states in \cite{primaatmaja2019almost} must modulate eight phases accurately, which introduces difficulties into the phase modulation system and may also limit the repetition rate of experiments. Using 4 intensities, our achievable distance is longer under the same conditions; however, the semi-definite programming method that was proposed by \cite{primaatmaja2019almost} is more novel and general.

%
In summary, we proposed a practically useful method that overcomes the disadvantages of NPP-TFQKD. By adding decoy mode 2, we obtain additional constraints that enable us to estimate $\overline{I}_{AE}$ tighter. The main strategy behind our method is that the inner products of the quantum states of Eve's system can be well estimated via the proposed protocol, whereas the previous protocol assumes that these inner products attain the worst possible value. This work improves the communication distance substantially. According to the result of the simulation, our method of 5 intensities in decoy mode 1 and 4 intensities in decoy mode 2 is close to the infinite decoy state of PM-QKD in terms of the communication distance, while the advantage of phase post-selection not being required is retained.

Experimentally, the manipulation of decoy mode 2 is similar to that of code mode and the random intensities are a subset of decoy mode 1. Thus, our modification does not introduce any additional difficulties into the experimental system.

Our method analyzes why the upper bound on the information leakage in \cite{cui2018phase} is too loose and provides a tighter bound such that the achievable distance reaches those of PM-QKD and \cite{lin2018simple}. We also propose the finite decoy state method, which renders the protocol useful in practice. 

Some problems remain to be solved, such as the finite size key effect and the protocol with imperfect devices.


\section{acknowledgment}
This work has been supported by the National Key Research and Development Program of China (Grant No. 2016YFA0302600), the National Natural Science Foundation of China (Grant Nos. 61822115, 61775207, 61702469, 61771439, 61622506, 61627820, 61575183), National Cryptography Development Fund (Grant No. MMJJ20170120) and Anhui Initiative in Quantum Information Technologies.

\appendix

\section{improved NPP-TFQKD with finite decoy-state}

In this section, we will describe how to estimate a tight bound of $I_{AE}$ with a finite decoy state in detail. The overline and underline denote upper and lower bounds, respectively. According to Ref. \cite{cui2018phase}, the key step of Eq.(\ref{I_EA_upper}) is the estimation of constraints on $x_{ee}$, $x_{eo}$, $x_{oe}$ and $x_{oo}$. In the original protocol, the constraints are specified as Eq.(\ref{cons}), which are too loose. By introducing decoy mode 2, tighter constraints can be obtained. Let's consider $x_{ee}$ as an example:

\begin{equation}
\begin{aligned}
\label{App_xee}
\underline{\Omega}_{ee}^{\mu\mu}+\underline{\Phi}_{ee}^{\mu\mu} \leqslant x_{ee}\leqslant \overline{\Omega}_{ee}^{\mu\mu}+\overline{\Phi}_{ee}^{\mu\mu},
\end{aligned}
\end{equation}
where 
\begin{equation}
\label{App_Omega_upper}
\overline{\Omega}_{ee}^{\mu\mu}  =  \sum_{n\in{\mathbb{E}},m\in{\mathbb{E}}}p^\mu_np^\mu_m\overline{Y}_{n,m},
\end{equation}
and
\begin{equation}
\label{App_Omega_lower}
\underline{\Omega}_{ee}^{\mu\mu}  =  \sum_{n\in{\mathbb{E}},m\in{\mathbb{E}}}p^\mu_np^\mu_m\underline{Y}_{n,m},
\end{equation}
where $\overline{Y}_{nm}$ in Eq.(\ref{App_Omega_upper})($\underline{Y}_{nm}$ in Eq.(\ref{App_Omega_lower}) ) can be estimated via linear programming, as shown in the appendix of Ref. \cite{cui2018phase}.

The bounds of ${\Phi}_{ee}^{\mu\mu}$ are estimated via linear programming. The upper bound is as follows:
\begin{equation}
\begin{aligned}
\label{App_linear_prog}
&\text{max}:\ \   {\Phi}_{ee}^{\mu\mu}=\sum_{n,k\in{\mathbb{E}},m,l\in{\mathbb{E}}}^{l>m}\sqrt{{p^{\mu}_{n}p^{\mu}_{m}p^{\mu}_{k}p^{\mu}_{l}}} y_{n,m,k,l}\\
&\text{s.t.}:\\
&Q_{\omega_1 \omega_2}^{c} - Q_{\omega_1 \omega_2}^{d} 
=\sum_{\mathbb{A},\mathbb{B}}^{\mathbb{A},\mathbb{B}\in{\{\mathbb{O},\mathbb{E}\}}}
\sum_{n,k\in{\mathbb{A}},m,l\in{\mathbb{B}}}^{l>m}
\sqrt{{p^{\omega_1}_{n}p^{\omega_2}_{m}p^{\omega_1}_{k}p^{\omega_2}_{l}}}y_{n,m,k,l},\\
&y_{n,m,k,l}\in{[-2\sqrt{\overline{Y}_{n,m}\overline{Y}_{k,l}},2\sqrt{\overline{Y}_{n,m}\overline{Y}_{k,l}}]},\\
&\omega_1 , \omega_2 \in \{\mu,\nu_1,\nu_2, o\},
\end{aligned}
\end{equation}
where $y_{n,m,k,l}$ is the variable of the objective function. By adjusting $y_{n,m,k,l}$ without violating the constraint functions, we obtain the maximum value of the objective function, which is the upper bound of ${\Phi}_{ee}^{\mu\mu}$.

Similarly, by modifying the objective function to

\begin{equation}
\begin{aligned}
\label{App_linear_prog_lower}
&\text{min:}\ \   {\Phi}_{ee}^{\mu\mu}=\sum_{n,k\in{\mathbb{E}},m,l\in{\mathbb{E}}}^{l>m}\sqrt{{p^{\mu}_{n}p^{\mu}_{m}p^{\mu}_{k}p^{\mu}_{l}}} y_{n,m,k,l},\\
\end{aligned}
\end{equation}
the lower bound of ${\Phi}_{ee}^{\mu\mu}$ can be estimated.

Since $\omega_1, \omega_2 \in \{\mu,\nu_1,\nu_2, o\}$, there are 16 constraints in the linear programming equation in total. By solving Eq.(\ref{App_linear_prog}), the upper bound of ${\Phi}_{ee}^{\mu\mu}$ is obtained. $\overline{\Phi}_{eo}^{\mu\mu}$, $\overline{\Phi}_{oe}^{\mu\mu}$ and $\overline{\Phi}_{oo}^{\mu\mu}$ can be obtained similarly.

After all $\overline{\Omega}$ and $\overline{\Phi}$ have been estimated, the constraints of Eq.(\ref{I_EA_upper}) is reexpressed as

\begin{equation}
\begin{aligned}
\label{App_cons}
&\underline{\Omega}_{ee}^{\mu\mu}+\underline{\Phi}_{ee}^{\mu\mu} \leqslant x_{ee}\leqslant \overline{\Omega}_{ee}^{\mu\mu}+\overline{\Phi}_{ee}^{\mu\mu},\\
&\underline{\Omega}_{eo}^{\mu\mu}+\underline{\Phi}_{oe}^{\mu\mu} \leqslant x_{oe}\leqslant \overline{\Omega}_{oe}^{\mu\mu}+\overline{\Phi}_{oe}^{\mu\mu},\\
&\underline{\Omega}_{eo}^{\mu\mu}+\underline{\Phi}_{eo}^{\mu\mu} \leqslant x_{eo}\leqslant \overline{\Omega}_{eo}^{\mu\mu}+\overline{\Phi}_{eo}^{\mu\mu},\\
&\underline{\Omega}_{oo}^{\mu\mu}+\underline{\Phi}_{oo}^{\mu\mu} \leqslant x_{oo}\leqslant \overline{\Omega}_{oo}^{\mu\mu}+\overline{\Phi}_{oo}^{\mu\mu},\\
&x_{ee}+x_{oe}+x_{oo}+x_{eo}=Q^c_{\mu\mu}.
\end{aligned}
\end{equation}

By applying the new constraints Eq.(\ref{App_cons}) in Eq.(\ref{I_EA_upper}), a tighter bound $\overline{I}_{AE}$ is obtained and improved performance of NPP-TFQKD can be expected.

\nocite{*}

\bibliography{apssamp}

\end{document}